\documentclass[10pt]{iopart}
\usepackage{graphicx,iopams}
\begin{document}
\jl{12}
\submitted

\note[Semi-empirical formulation of multiple scattering]{Semi-empirical formulation of multiple scattering for Gaussian beam model of heavy charged particles stopping in tissue-like matter}

\author{Nobuyuki Kanematsu}

\address{Department of Accelerator and Medical Physics, Research Center for Charged Particle Therapy, National Institute of Radiological Sciences, 4-9-1 Anagawa, Inage-ku, Chiba 263-8555, Japan}

\eads{nkanemat@nirs.go.jp}

\begin{abstract}
Dose calculation for radiotherapy with protons and heavier ions deals with a large volume of path integrals involving a scattering power of body tissue. This work provides a simple model for such demanding applications.
There is an approximate linearity between RMS end-point displacement and range of incident particles in water, empirically found in measurements and detailed calculations.
This fact was translated into a simple linear formula, from which the scattering power that is only inversely proportional to residual range was derived.
The simplicity enabled analytical formulation for ions stopping in water, which was designed to be equivalent with the extended Highland model and agreed with measurements within 2\% or 0.02 cm in RMS displacement.
The simplicity will also improve the efficiency of numerical path integrals in the presence of heterogeneity.
\end{abstract}


\section{Introduction}

The essence of radiotherapy with protons and heavier ions lies in precise control of incident particles that are designed to stop in tumor volume. 
The targeting precision will be inevitably deteriorated by multiple scattering in beam modifiers and patient body and such effects must be accurately handled for dose calculations in treatment planning.
On the other hand, simplicity and efficiency are also essential in clinical practice and there has always been need for a computational method that balances all these demanding and conflicting requirements.

Fermi and then Eyges (1948) developed a general theory for charged particles that undergo energy loss and multiple scattering in matter.
A group of particles is approximated as a Gaussian beam growing in space with statistical variances
\begin{eqnarray}
\overline{\theta^2}(x) &=& \overline{\theta^2}(0)+\int_0^x T(x')\, \rmd x', 
\label{eq:int0}
\\
\overline{y \theta}(x) &=& \overline{y \theta}(0)+\overline{\theta^2}(0)\,x+\int_0^x (x-x')\, T(x')\, \rmd x', 
\label{eq:int1}
\\
\overline{y^2}(x) &=& \overline{y^2}(0)+2\,\overline{y \theta}(0)\,x+\overline{\theta^2}(0)\,x^2+\int_0^x (x-x')^2\, T(x')\, \rmd x',
\label{eq:int2}
\end{eqnarray}
where $x$ is the longitudinal position and $y$ and $\theta$ are the projected particle position and angle.
The original Fermi-Eyges theory adopted purely Gaussian approximation (Rossi and Greisen 1948) with (projected) scattering power
\begin{equation}
T = \frac{E_\mathrm{s}^2}{X_0}\left(\frac{z}{p v}\right)^2,
\label{eq:tfr}
\end{equation}
where $E_\mathrm{s} = m_e\, c^2 \sqrt{2 \pi/\alpha} \approx 15.0$ MeV is a constant, $X_0$ is the radiation length of the material, and $z$, $p$, and $v$ are the charge, the momentum, and the velocity of the particles.
The Fermi--Rossi formula \eref{eq:tfr} totally ignores effects of large-angle single scattering (Hanson \etal 1951) and was found to be inaccurate (Wong \etal 1990).

Based on formulations by Highland (1975, 1979) and Gottschalk \etal (1993), Kanematsu (2008b) proposed a scattering power with correction for the single-scattering effect, although within the Gaussian approximation,
\begin{equation}
T = 0.970 \left(1+\frac{\ln\ell}{20.7}\right) \left(1+\frac{\ln\ell}{22.7}\right) \frac{E_\mathrm{s}^2}{X_0}\left(\frac{z}{p v}\right)^2,
\label{eq:thk}
\end{equation}
where $\ell = \int_0^x \rmd x'/X_0(x')$ is the radiative path length.
Although it would be difficult to calculate integrals \eref{eq:int0}--\eref{eq:int2} because of the embedded integral in the $\ell$ terms, Kanematsu (2008b) further derived an approximate formula for the RMS displacement of incident ions at the end point in homogeneous matter as
\begin{equation}
\fl
\sigma_{y 0}(R_0) = 
\frac{E_\mathrm{s}}{\rm MeV}
\frac{z^{1-\kappa}}{\sqrt{3-\kappa}}
\left(\frac{m}{m_p}\right)^{\frac{\kappa}{2}-1}
\sqrt{0.816-0.082\ln\frac{\rho_\mathrm{S} X_0}{R_0}}
\left(\frac{R_0}{\lambda\,{\rm cm}}\right)^{-\frac{\kappa}{2}}
\sqrt{\frac{R_0^3}{\rho_\mathrm{S}^3 X_0}},
\label{eq:sy0}
\end{equation}
where $m/m_p$ is the ion mass in units of the proton mass, $R_0$ is the expected in-water range on the incidence, $\rho_\mathrm{S}$ is the stopping-power ratio of the matter relative to water, and $\kappa = 1.08$ and $\lambda = 4.67 \times 10^{-4}$ are constants.

Despite the complex involvement of variable $R_0$ in \eref{eq:sy0}, Kanematsu (2008b) found the $\sigma_{y 0}$--$R_0$ relation for ions in water to be very linear. 
In fact, Preston and Kohler of Harvard Cyclotron Laboratory knew the linear relation and derived universal curve $\sigma_y(x)/\sigma_{y 0}(R_0) = \sqrt{3 x_R^2 -2 x_R -2(1-x_R)^2\ln(1-x_R)}$ with $x_R = \rho_\mathrm{S}\,x/R_0$ for relative growth of RMS displacement $\sigma_y = (\overline{y^2})^{1/2}$ in homogeneous matter in an unpublished work in 1968.
Starting with the empirical linear relation, this work is aimed to develop a simple and general multiple-scattering model to improve efficiency of numerical heterogeneity handling and to enable further analytical beam modeling.

\section{Materials and methods}

\subsection{Linear-displacement model}

Linear approximation $\sigma_{y 0} \propto R_0$ for homogeneous systems greatly simplifies \eref{eq:sy0} to
\begin{equation}
\sigma_{y 0} (R_0) = 
0.0224\, z^{-0.08} \left(\frac{m}{m_p}\right)^{-0.46}
\sqrt{\frac{{X_0}_\mathrm{w}}{\rho_\mathrm{S} X_0}} \,
\frac{R_0}{\rho_\mathrm{S}},
\label{eq:sy0l}
\end{equation}
where ${X_0}_\mathrm{w} = 36.08$ cm is the radiation length of water, $\sqrt{{X_0}_\mathrm{w}/(\rho_\mathrm{S} X_0)}$ is the scattering/stopping ratio of the material relative to water, and $R_0/\rho_\mathrm{S}$ is the geometrical range. 
\Eref{eq:sy0l} was calibrated to \eref{eq:sy0} for water at $R_0 = X_0$.

\subsection{Formulation of new scattering power}

Equation $\overline{y^2} = \sigma_{y 0}^2$ at the end point $x = R_0/\rho_\mathrm{S}$ associates \eref{eq:int2} and \eref{eq:sy0l} as
\begin{equation}
\int_0^{R_0} \frac{R^2}{\rho_\mathrm{S}^2} T \frac{\rmd R}{\rho_\mathrm{S}} = 0.0224^2\, z^{-0.16} \left(\frac{m}{m_p}\right)^{-0.92} 
\frac{{X_0}_\mathrm{w}}{\rho_\mathrm{S} X_0} 
\left(\frac{R_0}{\rho_\mathrm{S}}\right)^2,
\end{equation}
to lead to another scattering power
\begin{equation}
T = f_{m z}\,\frac{{X_0}_\mathrm{w}}{X_0} \frac{1}{R}, \qquad
f_{m z} = \left(1.00 \times 10^{-3}\right) z^{-0.16} \left(\frac{m}{m_p}\right)^{-0.92},
\label{eq:tr}
\end{equation}
where $f_{m z}$ is the particle-type-dependent factor.
The scattering power is inherently applicable to any heterogeneous system by numerical integral of \eref{eq:int0}--\eref{eq:int2}.

\subsection{Comparison of models and measurements}

We examined these Fermi-Rossi \eref{eq:tfr}, extended Highland \eref{eq:thk}, and linear-displacement \eref{eq:tr} models with unpublished measurements by Phillips (Hollmark \etal 2004), those by Preston and Kohler (Kanematsu 2008b), and Moli\`{e}re-Hanson calculations by Deasy (1998).
We took growths of RMS displacement $\sigma_y(x)$ with depth $x$ for $R_0 = 29.4$ cm protons, $R_0 = 29.4$ cm helium ions, and $R_0= 29.7$ cm carbon ions in water and RMS end-point displacements $\sigma_{y 0}(R_0)$ for them with varied incident range $R_0$.

\subsection{Formulas for homogeneous systems}

For a point mono-directional ion beam with in-water range $R_0$ incident into homogeneous matter with constant $\rho_\mathrm{S}$ and $X_0$, equations \eref{eq:int0}--\eref{eq:int2} are analytically integrated to
\begin{eqnarray}
\overline{\theta^2} = f_{m z}\, \frac{{X_0}_\mathrm{w}}{\rho_\mathrm{S} X_0} \ln\frac{R_0}{R},
\label{eq:int0h}
\\
\overline{y\theta} = f_{m z}\,\frac{{X_0}_\mathrm{w}}{\rho_\mathrm{S} X_0} \left(1-\frac{R}{R_0}+\frac{R}{R_0} \ln\frac{R}{R_0}\right) \frac{R_0}{\rho_\mathrm{S}},
\label{eq:int1h}
\\
\overline{y^2} = f_{m z}\,\frac{{X_0}_\mathrm{w}}{\rho_\mathrm{S} X_0}
\left(\frac{1}{2}+\frac{1}{2} \frac{R^2}{R_0^2}-2\frac{R}{R_0}-\frac{R^2}{R_0^2} \ln\frac{R}{R_0} \right) \left(\frac{R_0}{\rho_\mathrm{S}}\right)^2,
\label{eq:int2h}
\end{eqnarray}
as a function of residual range $R = R_0-\rho_\mathrm{S} x$ at distance $x$.
\Eref{eq:int2h} in fact reduces to the universal curve by Preston and Kohler.

A radiation field at a given $x$ position can be effectively or virtually modeled with a source, ignoring the matter (ICRU-35 1984). The effective extended source is at $x_\mathrm{e} = x-\overline{y\theta}(x)/\overline{\theta^2}(x)$, where $\overline{y^2}$ would be minimum in vacuum.
The virtual point source is at $x_\mathrm{v} = x-\overline{y^2}(x)/\overline{y\theta}(x)$, from which radiating particles would form a field of equivalent divergence.
Similarly, the effective scattering point is at $x_\mathrm{s} = x-[\overline{y^2}(x)/\overline{\theta^2}(x)]^{1/2}$, at which a point-like scattering would cause equivalent RMS angle and displacement (Gottschalk \etal 1993).

\section{Results}

\begin{figure}
\includegraphics{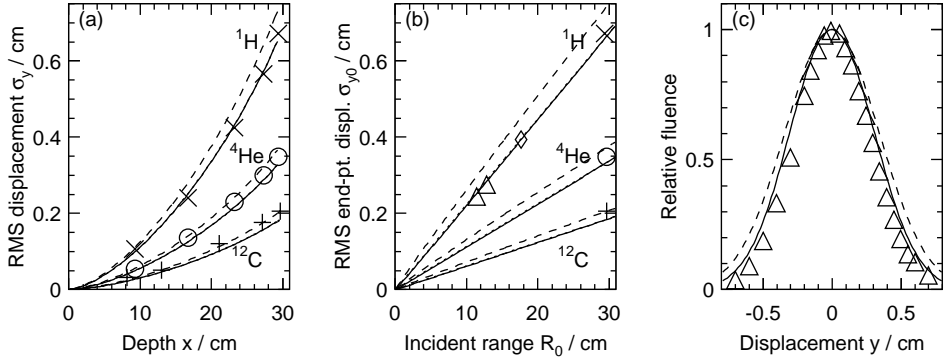}
\caption{(a) RMS displacement $\sigma_y$ with depth $x$ and (b) RMS end-point displacement $\sigma_{y 0}$ with incident range $R_0$ for protons, helium ions, and carbon ions in water, calculated with the Fermi-Rossi (dashed), extended Highland (solid), linear-displacement (dotted--overlapping with the solid) models along with a Moli\`{e}re-Hanson calculation by Deasy ($\diamond$) and measurements by Phillips ($\times$, $\circ$, $+$) and by Preston and Kohler ($\triangle$).
(c) End-point lateral shape of a proton pencil beam in water ($R_0 = 11.4$ cm) measured by Preston and Kohler ($\triangle$) and estimated by Fermi-Rossi (dashed) and the present (solid) formulations.}
\label{fig:displacements}
\end{figure}

\Fref{fig:displacements}(a) shows RMS-displacement growths $\sigma_y(x)$ and \fref{fig:displacements}(b) shows RMS end-point displacements $\sigma_{y 0}(R_0)$.
The present model was virtually identical to the extended Highland model with deviations from measurements or Moli\`{e}re-Hanson calculations within either 2\% or 0.02 cm, while the Fermi-Rossi model overestimated the RMS displacements by nearly 10\%.
\Fref{fig:displacements}(c) shows an end-point shape of a pencil beam measured by Preston and Kohler for $R_0 = 11.4$ cm protons in water and curves estimated by Fermi-Rossi and the present formulations with additional $(0.168~{\rm cm})^2$ to $\sigma_y^2$ for incident beam emittance in their experiment.
The Gaussian approximation was in fact adequate in this case.

\begin{figure}
\indented{\item[]\includegraphics{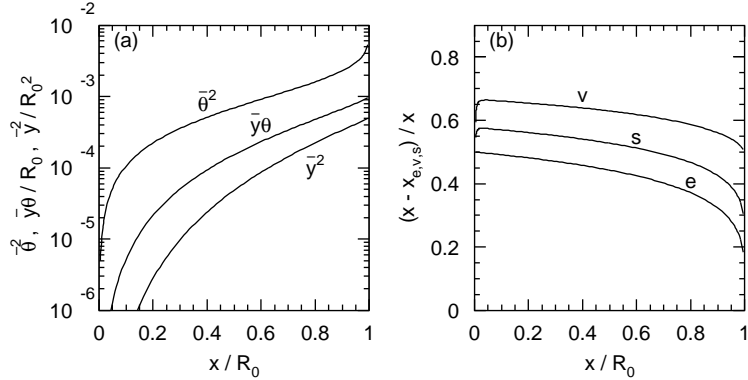}}
\caption{Analytical curves of (a) parameters $\overline{\theta^2}$, $\overline{y\theta}$, and $\overline{y^2}$ and (b) relative distances to effective extended source (e), the virtual point source (v), and effective scattering point (s) as a function of normalized depth $x/R_0$ for protons in water.}
\label{fig:sources}
\end{figure}

\Fref{fig:sources}(a) shows analytical variances $\overline{\theta^2}$, $\overline{y \theta}$, and $\overline{y^2}$ growing with depth $x$ for protons in water.
\Fref{fig:sources}(b) shows the relative distances from current position $x$ to the effective extended source $x_\mathrm{e}$, the virtual point source $x_\mathrm{v}$, and the effective scattering point $x_\mathrm{s}$.
They approach $(x-x_\mathrm{e})/x \to 1/2$, $(x-x_\mathrm{v})/x \to 2/3$, and $(x-x_\mathrm{s})/x \to 1/\sqrt{3}$ at the $x \to 0$ limit.
Increase of the scattering power with depth moves these points relatively closer to the current position.

\section{Discussion}

In application of Bragg peaks, the end-point displacement is the most important, for which the Gaussian approximation was valid in the proton experiment by Preston and Kohler. 
In fact, the single-scattering effect is theoretically small for a thick target (Hanson \etal 1951).
Although ions suffer nuclear interactions with resultant fragments that generally scatter at large angles (Matsufuji \etal 2005), their contributions may be relatively less significant at the Bragg peaks. 
The present formulation would be thus adequate for radiotherapy.

The linear-displacement model with the Fermi-Eyges theory has brought general formulas for ions, including the universal curve intuitively derived by Preston and Kohler without explicit formulation of the scattering power.
In the present model, the kinematic properties are encapsulated in residual range $R = R_0 - \int_0^x \rho_\mathrm{S} \rmd x'$ that is always tracked in beam transport.
The ion-type dependence in scattering angle and displacement is simply as proportional to $\sqrt{f_{m z}}$, which leads to 50\% for $^4$He, 28\% for $^{12}$C, and 24\% for $^{16}$O with respect to that of protons for a given incident range. 
These numbers coincide with detailed numerical calculations by Hollmark \etal (2004).

\begin{figure}
\indented{\item[]\includegraphics{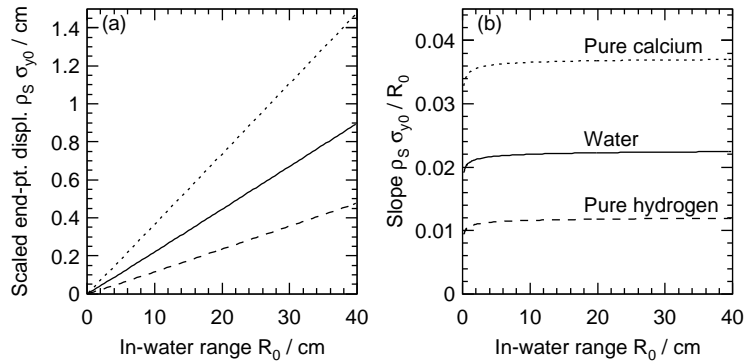}}
\caption{Scaled RMS end-point displacement $\rho_\mathrm{S} \sigma_{y 0}$ as a function of incident in-water range $R_0$ and its slope $\rho_\mathrm{S} \sigma_{y 0}/R_0$ for water (solid), pure hydrogen (dashed), and pure calcium (dotted) calculated with the extended Highland model.}
\label{fig:linearity}
\end{figure}

The linearity between end-point displacement and range observed for water is the basis of the present model.
Its validity for general body-tissue materials is not obvious.
We here examine water and two extreme elements hydrogen ($\rho_\mathrm{S} X_0 \approx 113$ cm) and calcium ($\rho_\mathrm{S} X_0 \approx 14.5$ cm) among major elements of body tissues (ICRU-46 1992), using \eref{eq:sy0} in the extended Highland model.
\Fref{fig:linearity} shows their $R_0$--$\sigma_{y 0}$ relations with geometrical scale correction and indicates that the linearity will generally hold for body-tissue elements.
However, the elemental linearity may not truly warrant the validity for systems heterogeneous in atomic compositions.
The scattering power \eref{eq:tr} only depends on the residual range that is irrelevant to the multiple scattering accumulated in the other upstream materials, whereas the accumulation should influence the single-scattering effect (Kanematsu 2008b). 
In other words, the present model implicitly assumes heterogeneity in density only.

In the current practice of treatment planning, the patient heterogeneity is normally modeled with variable-density water (Kanematsu \etal 2003), for which the present model is rigorous with further simplified scattering power
\begin{eqnarray}
T = f_{m z}\,\frac{\rho_\mathrm{S}}{R}.
\label{eq:tw}
\end{eqnarray}
The simplicity will minimize computation of the integrands in path integrals \eref{eq:int0}--\eref{eq:int2} for demanding dose calculations (Kanematsu \etal 1998, 2006, 2008a).

\section{Conclusions}

A novel multiple-scattering model has been formulated based on the fact such that the RMS end-point displacement is proportional to the incident range in water.
The model was designed to be equivalent with the extended Highland model for stopping ions in water and agreed with measurements within 2\% or 0.02 cm in RMS displacement.

The resultant scattering-power formula that is only inversely proportional to residual range is much simpler than former formulations and can be used in the framework of the Fermi-Eyges theory for Gaussian-beam transport in tissue-like matter.
The simplicity enables analytical beam modeling for homogeneous systems and improves efficiency of numerical path integrals for heterogeneous systems. 
The present model is ideal for demanding dose calculations in treatment planning of heavy-charged-particle radiotherapy.

\References

\item[]
Deasy J O 1998 A proton dose calculation algorithm for conformal therapy simulations based on Moli\`ere's theory of lateral deflections {\it Med. Phys.} {\bf 25} 476--83

\item[]
Eyges L 1948 Multiple scattering with energy loss {\it Phys. Rev.} {\bf 74} 1534--5

\item[]
Gottschalk B, Koehler A M, Schneider R J, Sisterson J M and Wagner M S 1993 Multiple Coublomb scattering of 160 MeV protons {\it Nucl. Instrum. Methods} B {\bf 74} 467--90

\item[]
Hanson A O, Lanzl L H, Lyman E M and Scott M B 1951 Measurement of multiple scattering of 15.7-MeV electrons {\it Phys. Rev.} {\bf 84} 634--7

\item[]
 Highland V L 1975 Some practical remarks on multiple scattering {\it Nucl. Instrum. Methods} {\bf 129} 497--9
\dash 1979 (erratum) {\it Nucl. Instrum. Methods} {\bf 161} 171
 
\item[]
Hollmark M, Uhrdin J, D\v{z} B, Gudowska I and Brahme A 2004 Influence of multiple scattering and energy loss straggling on the absorbed dose distributions of therapeutic light ion beams: I. Analytical pencil beam model \PMB {\bf 49} 3247--65

\item[]
ICRU-35 1984 Radiation dosimetry: electron beams with energies between 1 and 50 MeV {\it ICRU Report} 35 (Bethesda, MD: ICRU)

\item[]
ICRU-46 1992 Photon, electron, proton and neutron interaction data for body tissues {\it ICRU Report} 46 (Bethesda, MD: ICRU) 

\item[]
Kanematsu N, Akagi T, Futami Y, Higashi A, Kanai T, Matsufuji N, Tomura H and Yamashita H 1998 A proton dose calculation code for treatment planning based on the pencil beam algorithm {\it Jpn. J. Med. Phys.} {\bf 18} 88--103

\item[]
Kanematsu N, Matsufuji N, Kohno R, Minohara S and Kanai T 2003 A CT calibration method based on the polybinary tissue model for radiotherapy treatment planning \PMB {\bf 48} 1053--64

\item[]
Kanematsu N, Akagi T, Takatani Y, Yonai S, Sakamoto H and Yamashita H 2006 Extended collimator model for pencil-beam dose calculation in proton radiotherapy \PMB {\bf 51} 4807--17

\item[]
Kanematsu N, Yonai S and Ishizaki A 2008a The grid-dose-spreading algorithm for dose distribution calculation in heavy charged particle radiotherapy {\it Med. Phys.} {\bf 35} 602--8

\item[]
Kanematsu N 2008b Alternative scattering power for Gaussian beam model of heavy charged particles {\it Nucl. Instrum. Methods} B {\bf 266} 5056--62

\item[]
Matsufuji N, Komori M, Sasaki H, Akiu K, Ogawa M, Fukumura A, Urakabe E, Inaniwa T, Nishio T, Kohno T and Kanai T 2005 Spatial fragment distribution from a therapeutic pencil-like carbon beam in water \PMB {\bf 50} 3393--403

\item[]
Rossi B and Greisen K 1941 Cosmic ray theory {\it Rev. Mod. Phys.} {\bf 13} 240--309

\item[]
Wong M, Schimmerling W, Phillips M H, Ledewigt B A, Landis D A, Walton J T and Curtis S B 1990 {\it Med. Phys.} {\bf 17} 163--71

\endrefs

\end{document}